\documentclass{ws-ijmpcs}

\newcommand{\be}{\begin{equation}}
\newcommand{\ee}{\end{equation}}
\newcommand{\bea}{\begin{eqnarray}}
\newcommand{\eea}{\end{eqnarray}}

\begin{document}

\markboth{Hyun Min Lee}
{Reheating the Universe at Criticality}

\catchline{}{}{}{}{}
%
%%%%%%%%%%%%%%%%%%%%%%%%%%%%%%%%%%%%%%%%%%%%%%%%%%%%%%%%%%%%%%%%%%%%

\title{Reheating the Universe at Criticality}

\author{Hyun Min Lee}

\address{Department of Physics, Chung-Ang University, 06974 Seoul, Korea.\\
hminlee@can.ac.kr \\ Prepared for the proceedings of CosPA 2015, \\ IBS-CTPU, Daejeon, Korea, 12-16 October, 2015.}

%\author{Second Author}

%\address{Group, Laboratory, Address\\
%City, State ZIP/Zone, Country\\
%second\_author@domain\_name}

\maketitle

%\begin{history}
%\received{Day Month Year}
%\revised{Day Month Year}
%\published{Day Month Year}
%\end{history}

\begin{abstract}
We present the general discussion on the inflection point inflation with small or large inflaton fields and show the effects of reheating dynamics on the inflationary predictions. In order to compare the model predictions with precisely measured CMB anisotropies and constrain the inflation models, the knowledge of the reheating dynamics is required. Inflection point inflation extended to the trans-Planckian regime can accommodate a sizable tensor-to-scalar ratio at the detectable level in the future CMB experiments.    
\end{abstract}

\ccode{PACS numbers:98.80.-k, 98.80.Cq, 04.20.-q.}

\section{Introduction}	

The measurement of Cosmics Microwave Background (CMB) anisotropies in Planck (TT$+$low-$l$ polarization)\cite{planck} shows the spectral index, $n_s=0.9652\pm 0.0047$ at $95\%$ C.L. and the bound on the tensor-to-scalar ratio at $k=0.002\,{\rm Mpc}^{-1}$ as $r<0.10$ with no running and $r<0.18$ with running at $95\%$ C.L.    No large non-Gaussianity is observed, for instance, a local-shape non-Gaussianity is constrained to $f^{\rm local}_{NL}=0.8\pm 5.0$ from T$+$E at $68\%$ C.L.\cite{planck}, so canonical single field inflations are favored.
On the other hand, there was an excitement due to the hint for primordial tensor mode from BICEP2, but there is no more significant excess after the dust estimated from Planck dust data at $353\,{\rm GHz}$ is subtracted, although the posterior probability function peaks at $r=0.05$\cite{tensor}. BiCEP2$+$ Keck array data at $150\,{\rm GHz}$ are consistent with Planck maps at higher frequencies\cite{tensor}, leading to the combined bound on the tensor-to-scalar ratio at $k=0.05\,{\rm Mpc}^{-1}$ as $r<0.12$ at $95\%$ C.L.

%models and predictions
A lot of inflation models have been suggested, ranging from small field cases such as hiltop inflation to large field cases such as chaotic inflation. Chaotic inflation models with trans-Planckian inflation field values leads to a sizable tensor-to-scalar ratio, so some of them such as quadratic and quartic inflations have been ruled out at more than $95\%$ C.L. But, in order to solve the horizon problem at the time of recombination, we need to know the history of the Universe from inflation period to radiation-dominated era down to the present in the Standard Big Bang Cosmology (SBBC). Since reheating dynamics is unknown, caution is needed to derive definite predictions of any model such as for spectral index and tensor-to-scalar ratio. 

In this review article, we review on the reheating dynamics and its effects on the inflationary predictions and discuss them in the context of inflection point inflations\cite{inflection}. We also briefly remark on the implications of the reheating effects on the running inflation with non-minimal coupling at criticality.

%reheating and inflection point inflation

\section{Reheating and number of efoldings}

The end of inflation is followed by reheating, during which the Universe is heated to a sufficiently high temperature to populate the SM particles at least for Big Bang Nucleosynthesis. Reheating dynamics can be parametrized by the equation of state for the inflaton field $w$ and the reheating temperature $T_{\rm rh}$. The equation of state $w$ is restricted to $-\frac{1}{3}\leq w \leq 1$ for ${\ddot a}\leq 0$, while the reheating temperature is in the range between $T_{\rm BBN}\sim 1-10\,{\rm MeV}$ and $T_{\rm max}=(45\pi^2 V_{\rm end} /g_*)^{1/4}$ with $H_{\rm end}$ being the Hubble scale at the end of inflation.

\begin{figure}[pt]
\centerline{\includegraphics[width=9cm]{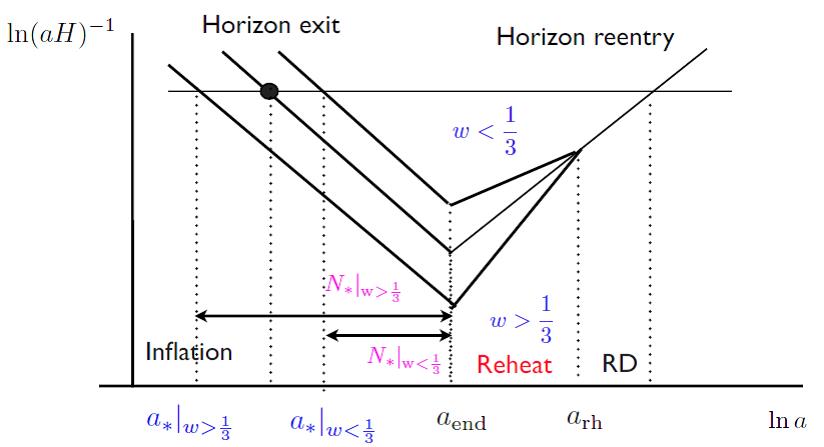}}
\vspace*{8pt}
\caption{Horizon distance as a function of scale factor (in log scales). The horizontal line corresponds to the scale of our interest entering the horizon during radiation.  The black dot is the moment that the scale of our interest exits the horizon. 
\label{RH}}
\end{figure}

The inflation period or the number of efoldings $N_*$ is identified between the horizon exit of cosmological scale and the end of slow-roll inflation. Then, slow-roll parameters during inflation depend on the form of inflaton potential and $N_*$. But, if reheating is not instantaneous, the growth of the horizon radius after inflation depends on the duration of reheating and the equation of state during reheating. As the Universe expands, the horizon radius decrease as $\ln(aH)^{-1}_I\sim -\ln a$ during inflation, whereas it increases as $\ln(aH)^{-1}_{\rm RD}\sim \ln a$ during radiation domination. On the other hand, the horizon radius behaves as $\ln(aH)^{-1}_{\rm rh}\sim \frac{1}{2}(1+3w)\ln a$ during reheating. As shown in Fig.~\ref{RH}, as compared to the case of instantaneous reheating, for $w<\frac{1}{3}$, the horizon radius at late times becomes smaller; for $w>\frac{1}{3}$, the horizon time scale at late times becomes larger. Therefore, the number of efoldings required for solving the horizon problem is smaller (larger) in the former (latter) cases, namely, $N_*|_{w>1/3}>N_*|_{w=1/3}>N_*|_{w<1/3}$.

Including the reheating effects, the number of efoldings is given \cite{inflection} by
\be
N_* = 61.4+\frac{3w-1}{12(1+w)}\,\ln\left(\frac{45}{\pi^2}\frac{V_*}{g_*(T_{\rm rh}^4)} \right)-\ln \left(\frac{V^{1/4}_*}{H_*} \right)
\ee
where $V_*(H_*)$ are the inflaton potential (Hubble scale) at horizon exit and $g_*(T_{\rm rh})$ is the number of effective relativistic degrees of freedom in thermal bath after reheating.

\section{Inflection point inflation}

\begin{figure}[pt]
\centerline{\includegraphics[width=8cm]{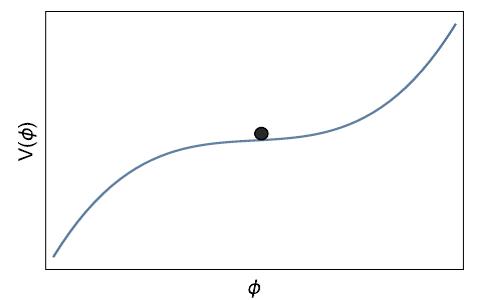}}
\vspace*{8pt}
\caption{Inflaton potential in inflection point inflation.  
\label{inflection}}
\end{figure}

In the case of small or large field inflation models, typically there are problems of graceful exit or quantum gravity corrections, respectively. 
On the other hand, in the case of inflection point inflation, the inflaton potential shown in Fig.~\ref{inflection} has a zero curvature and a small slope at certain field values smaller than Planck scale, so there is no such problem encountered in small or large field inflation models.
However, there must be a symmetry protecting the potential against quantum corrections at inflection point and the on-set of inflation must be free from initial condition problems in order not to reintroduce a fine-tuning problem.

Assuming that it is sufficient to Taylor expand the inflaton potential near inflection point $\phi_0$  up to cubic term as follows,
\be
 V=V_0 +\lambda_1 (\phi-\phi_0) + \frac{1}{3!} (\phi-\phi_0)^3
\ee
where $V_0, \lambda_1\, \lambda_3$ are constant parameters. 
We note that a positive $\lambda_1$ ensures graceful exit. 
Then, taking the constant term $V_0$ to be dominant for slow-roll inflation, we get slow-roll parameters as
\bea
\epsilon&=& \frac{1}{2}\left(\frac{\lambda_1+\frac{1}{2}\lambda_3(\phi-\phi_0)^2}{V_0} \right)^2, \\
\eta &=& \frac{\lambda_3}{V_0}\, (\phi-\phi_0).
\eea
On the other hand, the number of efoldings during inflation is given by
\be
N_*= \frac{N_{\rm max}}{\pi}\,\Big(\arctan\Big(\frac{N_{\rm max}}{2\pi}\,\eta_*\Big)-\arctan\Big(\frac{N_{\rm max}}{2\pi}\,\eta_{\rm end} \Big)  \Big) \label{efold}
\ee
where $N_{\rm max}\equiv \pi V_0\sqrt{2/(\lambda_1\lambda_3)}$ and $\eta_*$ is the $\eta$ parameter evaluated at horizon exit. 
For $\eta_{\rm end}=-1$, that is the case with small-field inflation,  $N_*\lesssim N_{\rm max}/2 $, so we need $V_0/\sqrt{\lambda_1\lambda_3}\ll 1$ for a sufficient number of efoldings $N_*$. On the other hand, for $|\eta_*|\lesssim|\eta_{\rm end}|\ll 1$, which is the case with large-field inflation, we get $N_*\ll N_{\rm max}/2$.

Consequently, the spectral index and the tensor-to-scalar ratio are given by
\bea
n_s &=& 1-3\left(\frac{\lambda_1}{V_0} \right)^2 +\frac{4\pi}{N_{\rm max}}\, \tan \left(\pi\,\frac{N_*}{N_{\rm max}}+\arctan\left(\frac{N_{\rm max}}{2\pi}\,\eta_{\rm end} \right) \right), \label{ns} \\
r&=& 8 \left(\frac{\lambda_1}{V_0} \right)^2.
\eea
This generalizes the results of Ref.\cite{dbrane} by including the case of large fields. 
Then, in order to satisfy the slow-roll conditions and the bound on the tensor-to-scalar ratio, we need to take $\lambda_1\ll V_0$. 
The CMB normalization, $A_s=\frac{1}{24\pi^2}\frac{V_*}{\epsilon_*}=2.196\times 10^{-9}$ at $k=0.05\,{\rm Mpc}^{-1}$, fixes one of the parameters as 
\be
V_0= 3.25\times 10^{-9} \left(\frac{\lambda_1}{3.63\times 10^{-10}} \right)^{2/3}. \label{p1}
\ee
On the other hand, the remaining parameters determine the tensor-to-scalar ratio and $N_{\rm max}$, respectively, as follows,
\bea
r &=& 0.10 \left(\frac{\lambda_1}{3.63\times 10^{-10}} \right)^{2/3}, \label{p2}\\
N_{\rm max} &=& 120 \left(\frac{3.98\times 10^{-11}}{\lambda_3} \right)^{1/2} \left(\frac{\lambda_1}{3.63\times 10^{-10}} \right)^{1/6}. \label{p3}
\eea

By using eqs.~(\ref{p1}) and (\ref{p3}), it can be shown that whether the inflaton field values during inflation are small or large can be determined by the ratio of parameters in the following\cite{inflection},
\be
\frac{V_0}{\lambda_3}= \left(\frac{N_{\rm max}}{120} \right)^2 \Big(\frac{r}{1.50\times 10^{-5}} \Big)^{1/2}. \label{discrim}
\ee
Thus, for a sub-Planckian field excursion with $|\phi_{\rm end}-\phi_0|\lesssim 1$ during inflation, we obtain a hierarchy between parameters, $\lambda_1\ll V_0\lesssim \lambda_3$, which results in a vey small tensor mode, $r\lesssim 10^{-5}$, for $N_{\rm max}\sim 120$. 
On the other hand, for a Planckian field excursion with $|\phi_{\rm end}-\phi_0|\gtrsim 1$, we need $\lambda_3\ll \lambda_1\ll V_0$. In this case, when $V_0/\lambda_3\gg 1$, the tensor-to-scalar ratio can be at the detectable level in the future CMB experiments such as COrE and LiteBIRD\cite{bucher,core,litebird}. 

We remark the reheating effects on the parameter space of the inflection point inflation.
When the equation of state satisfies $w<\frac{1}{3}$, the number of efoldings required for solving the horizon problem is smaller, so is $N_{\rm max}$ from eq.~(\ref{efold}).  The situation is the opposite for $w>\frac{1}{3}$.
For small-field inflation, the spectral index is sensitive to the change in $N_{\rm max}$ or the reheating temperature. Moreover, in this case, the tensor mode depends strongly on the reheating dynamics, as $\lambda_1$ changes much with $V_0/\lambda_3\lesssim 1$ remaining  in eqs.~(\ref{p2}) and (\ref{discrim}). 
On the other hand, for large-field inflation, for $V_0/\lambda_3\gtrsim 1$, the reheating effects are similar to the case in small-field inflation; for $V_0/\lambda_3\gg 1$, we can have a very large $N_{\rm max}$ and the $\eta$ contribution to the spectral index is small from eq.~(\ref{ns}), so the spectral index is insensitive to the reheating dynamics. 

The implications of the above general discussion on the concrete models of inflection point inflation have been discussed in Ref.~\cite{inflection}. In particular, when inflation with non-minimal coupling takes place near criticality where both the running quartic coupling for inflaton and its one-loop beta function are vanishingly small, the effective parameters in inflection point inflation can be obtained in terms of the non-minimal coupling, the inflaton field value at the inflection point and the two-loop beta function for the inflaton quartic coupling. The SM Higgs inflation and the $B-L$ Higgs inflation are such examples. In the latter case, when the $B-L$ Higgs mass is smaller than right-handed neutrino and $B-L$ gauge boson masses but it is heavier than twice the SM Higgs mass, the reheating temperature depends on the mixing quartic coupling between the $B-L$ and SM Higgs bosons. In the case with a small mixing quartic coupling, for instance, when the Higgs mass parameter is solely from the mixing quartic coupling after the $B-L$ is broken at a high scale, a very low reheating temperature can be achieved. Although the $B-L$ parameters can be limited under the condition of inflection point inflation and the collider bounds, the reheating dynamics must be also understood for the precise determination of the reheating temperature\cite{inflection}.

\section{Conclusions}
We have given a general discussion of the inflection point inflation and have shown how the effective parameters of the model are constrained by the CMB measurements. 
We discussed the effects of reheating dynamics on the inflationary predictions of the model in terms of the reheating temperature and the equation of state during reheating.  When inflection point inflation is realized by the quartic coupling running from low energies, the direct production of the inflaton at colliders can give a useful information for the inflaton couplings to the SM particles and the resultant reheating temperature.

\section*{Acknowledgments}

The work of HML is supported in part by Basic Science Research Program through the National Research Foundation of Korea (NRF) funded by the Ministry of Education, Science and Technology (2013R1A1A2007919).

%\appendix

%\section{Appendices}

%\begin{thebibliography}{000} %for 3 digits
%\begin{thebibliography}{00}  %for 2 digits

\end{document}